\documentclass{article}
\usepackage{frascatiphys,here,graphicx,subfigure}
\usepackage{amsmath}
\begin{document}
\title{ 
Search for $f_1(1285) \rightarrow \pi^+\pi^-\pi^0$ decay
with VES detector.
}
\author{
V.Dorofeev,
R.Dzheliadin,
A.Ekimov,
Yu.Gavrilov, \\
Yu.Gouz,
A.Ivashin,
V.Kabachenko,
I.Kachaev,  \\
A.Karyukhin,
Yu.Khokhlov,
V.Konstantinov,
M.Makouski, \\
V.Matveev,
A.Myagkov,
V.Nikolaenko,
A.Ostankov,
B.Polyakov,  \\
D.Ryabchikov,
N.Shalanda,
M.Soldatov,
A.A.Solodkov,
A.V.Solodkov,  \\
O.Solovianov, 
A.Zaitsev     \\
{\em IHEP, 142281, Protvino, Moscow region, Russia
} \\  \\
presented by V.Nikolaenko \\
}
\maketitle
\baselineskip=11.6pt
\begin{abstract}
The isospin violating decay $f_1(1285)\rightarrow\pi^+\pi^-\pi^0$ 
has been studied at VES facility.
This study is based at the statistics 
acquired 
in $\pi^- Be$ interactions
at 27, 36.6 and 41 GeV/c
in diffractive reaction
$\pi^- N \rightarrow (f_1 \pi^-) N$. 
The $f_1(1285) \rightarrow \pi^+\pi^-\pi^0$ decay
is observed. The ratio of decay probabilities
$BR(f_1(1285) \rightarrow \pi^+\pi^-\pi^0)$ to 
$BR(f_1(1285) \rightarrow \eta \pi^+\pi^-) \cdot
BR(\eta \rightarrow \gamma\gamma)$ 
is 
$\sim\:1.4\%$.   
\end{abstract}
\baselineskip=14pt

\section{Introduction.}


The decay  $f_1(1285) \rightarrow \pi^+\pi^-\pi^0$
violates the 
isospin symmetry.
It can proceed by means of $f_1(1285) \rightarrow a_1(1260)$
mixing and by a direct decay $f_1(1285) \rightarrow (\pi^+\pi^-\pi^0)$.
The $f_1(1285) \rightarrow a_1(1260)$ mixing is driven mainly by
the difference of light  quark mass $\Delta m = m_d - m_u$
 \cite{ref:coleman,ref:weinberg}.
Namely this $\Delta m$ is responsible for known decays
$\omega \rightarrow \pi^+\pi^-$, $\phi(1020)\rightarrow \pi^+\pi^-$,
$\eta \rightarrow 3\pi$ and $\eta' \rightarrow 3\pi$.
In the case of $f_1 \leftrightarrow a_1$ mixing it leads to 
$a_1$-like final states: $(\rho\pi),(f_0(600)\pi)$.
Another effect can contribute to the decay 
$f_1(1285) \rightarrow \pi^+\pi^-\pi^0$, namely 
the $a_0(980) \leftrightarrow f_0(980)$ mixing
predicted in 1979 \cite{ref:achasov1979}. 
Qualitatively speaking, loops with virtual
$K^+K^-$ and $K^0\bar K^0$ pairs cancel one another, but this cancellation is not perfect
due to the 
difference in mass of charged and neutral kaons.
The isospin symmetry violation reaches the maximum at the region between 
thresholds for pairs of charged and neutral kaons.
The amplitude of the isospin
violating transition depends on the couplings 
of scalar mesons
with $K\bar K$ pairs, in other words, it can shed light 
on the structure of scalars.
This phenomenon was discussed in details and 
several possibilities for its 
experimental observation were proposed, 
including a special
polarization experiment \cite{ref:achasov2004}, $f_1(1285)$ decays
\cite{ref:achasov1981}
and $J/\psi$ decays \cite{ref:wu2007}.  
Theoretial aspects of the expected 
$a_0(980) \leftrightarrow f_0(980)$ mixing
are discussed in details in recent paper 
\cite{ref:Hanhart2007}.


   Diffractive reaction $\pi^-N\rightarrow (f_1\pi^-)N \rightarrow
(\eta\pi^+\pi^-)\pi^-N$ represents a reach source of the $f_1(1285)$
mesons at low background. The branching ratio of  
$f_1\rightarrow a_0\pi$ decay is large, $BR=0.36\pm 0.07$ \cite{ref:PDGf1}.
The  process chain
\begin{equation}
f_1(1285) \rightarrow a_0(980)\pi^0 \rightarrow f_0(980)\pi^0
           \rightarrow (\pi^+\pi^-)\pi^0;
\end{equation}
is well suitable for a search of expected isospin violation.

\begin{figure}
  \includegraphics[height=.3\textheight]
{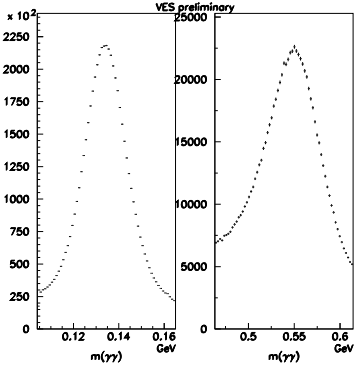}
  \caption{
Effective mass of $(\gamma\gamma$ pairs 
a) reaction $\pi^-N \rightarrow \pi^+\pi^-\pi^-\pi^0 N $;
b) reaction $\pi^-N \rightarrow \eta \pi^+\pi^-\pi^- N $;
}
\end{figure}
\section{Experimental procedure.}

   This study is based on the statistics acquired by the VES
experiment \cite{ref:VESEXP} in  interactions of a $\pi^-$ beam at 
the momentum of 27, 36.6 and  $GeV/c$ on a $Be$
target, in reaction 
\begin{equation}
\pi^-N\rightarrow\pi^+\pi^-\pi^-\pi^0 N.
\end{equation}
VES is a wide-aperture magnetic spectrometer equipped with a lead-glass
electromagnetic calorimeter and Cherenkov detectors for charged particle
identification. 
Events from reaction 
\begin{equation}
\pi^-N\rightarrow\pi^+\pi^-\pi^-\eta  N
\end{equation}
were selected also and used for normalization.
The $\pi^0$ and $\eta$ mesons were detected in the
$\gamma\gamma$  mode.
Selection criteria which have been applied for the selection
of the $(\pi^+\pi^-\pi^-\eta)$ events are described in 
\cite{ref:VES2006}. Similar selection procedure was used 
for the $(\pi^+\pi^-\pi^-\pi^0)$ events;
here the effective mass of two photons 
was requested in the range $(0.105,0.165)\:GeV/c^2$ (see Fig.1).
A kinematical fit to the $\eta$ or $\pi^0$ mass has been
performed, respectively.
%
%
%
\begin{figure}
  \includegraphics[height=.3\textheight]
{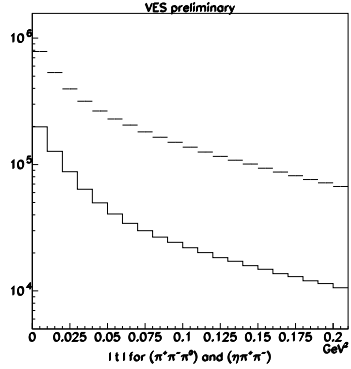}
  \caption{
$ |t'|$-distributions for reactions 
$\pi^-N \rightarrow \pi^+\pi^-\pi^-\pi^0 N $ ( upper distribution) and
$\pi^-N \rightarrow \eta \pi^+\pi^-\pi^- N $.
}
\end{figure}
%
%
%
%
The $t$-distributions for the reactions (2) and (3)
are shown in Fig. 2. 
The low $|t|$
region is relatively higher for the reaction (3),
which is a consequence of the diffractive production.

\begin{figure}
  \includegraphics[height=.3\textheight]
{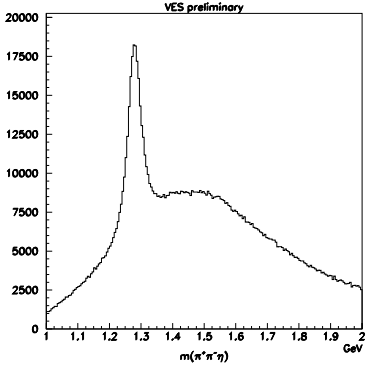}
  \caption{
Effective mass of $(\eta\pi^+\pi^-)$ system  produced in the reaction 
$\pi^- N \rightarrow (\eta\pi^+\pi^-\pi^-) N$
at low $t'$, $|t'|<0.04$ $GeV^2$. There are two entries per event.
}
\end{figure}
Fig.3 
demonstrates the 
$f_1(1285)$ signal
which is observed in the dominant 
decay channel,
$f_1 \rightarrow \eta\pi^+\pi^- \rightarrow \gamma\gamma\pi^+\pi^-$
at low momentum transfer region, $|t'|<0.04$. 
The estimated number of events in the
$f_1$ peak is $N_{\eta}=117600\pm 1300$, assuming the Breit-Wigner 
shape of the signal.
Concerning the $f_1$ production process, 
the results of the partial wave analysis of
$\eta\pi^+\pi^-\pi^-$ system \cite{ref:gouz1992} show  that 
the $(f_1\pi^-)$ system is produced in diffractive reaction.
The dominant wave  is $J^{PC}M\eta = 1^{++}0+$,
here $M$ is 
the spin 
projection and the $\eta$ denotes the exchange naturality.
Then the intermediate system with spin-parity $1^+$ decays
into $f_1(1285)$ and extra $\pi^-$, this is a $P$-wave decay.
Then the $f_1(1285)$ decays into $\eta\pi^+\pi^-$, this decay
also  includes a        $P$-wave.
The dominant angular term in the 
effective amplitude (which describes the chain of processes) is 
\begin{equation}
A \sim sin(\theta_1) \cdot \sin(\theta_2) \cdot sin(\phi_0 - \phi_2)
\end{equation}
here $\theta_1$ is the Gottfried-Jackson angle of the extra $\pi^-$; 
$\theta_2$ is polar 
angle of $\pi^0$ at the $f_1$ rest frame
with Z-axis going along the direction of the extra $\pi^-$
(so called "canonic system");
$\phi_0$ and $\phi_2$ are the azimuthal 
angles of the beam particle
and the $\pi^0$ at the same system.
The validity of this formula is demonstrated in Fig.4. 
Apart from the mass spectrum presented in Fig.3, 
similar 
distributions were accumulated 
in several intervals on the 
angular weight 
$W = |A|^2$.
The ratio of two mass spectra,
one of them for events at high $W$
and another one  
at low $W$, is shown.   
One can see that the angular weight $W$
strengthens the $f_1$ signal. 
This weight was  used      for the identification 
of the $f_1 \rightarrow \pi^+\pi^-\pi^0$ decay.

\begin{figure}
  \includegraphics[height=.3\textheight]
{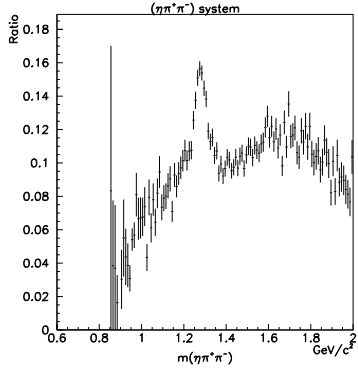}
  \caption{
Ratio of two $m(\eta\pi^+ \pi^-)$ spectra.
The distribution for events at $W>0.8$
is divided by the spectrum for events at  $W<0.2$ (see text).
}
\end{figure}

\begin{figure}
  \includegraphics[height=.42\textheight]
{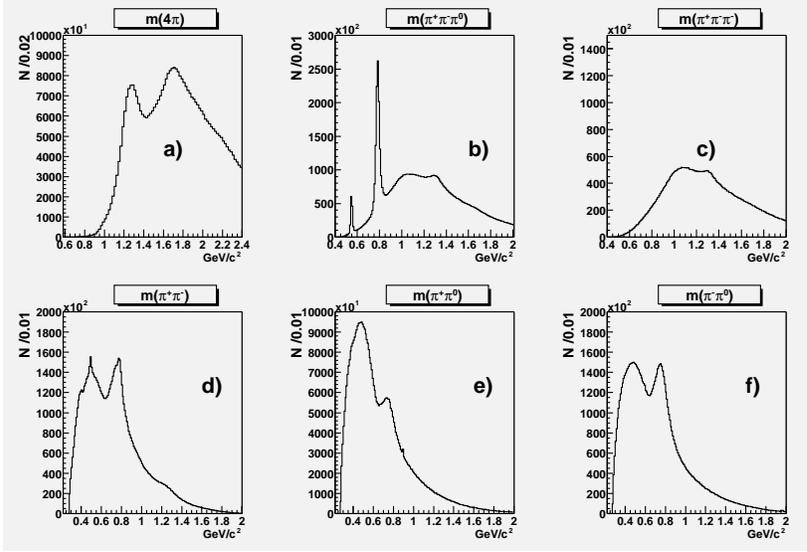}
  \caption{
Effective masses for $(\pi^+\pi^-\pi^-\pi^0)$ system.
a) total mass; b) $m(\pi^+\pi^-\pi^0)$ ; 
c) $m(\pi^+\pi^-\pi^-)$ ;
d) $m(\pi^+\pi^-)$, a zoom of the mass region 
from 680 to 880 $MeV$  is shown ;
e) $m(\pi^+\pi^0)$ ;
f) $m(\pi^-\pi^0)$ .
}
\end{figure}
Now we consider the general characteristics of the
reaction (2).
Fig.5 
demonstrates the mass spectra for the selected 
$(\pi^+\pi^-\pi^-\pi^0)$ sample. 
The $b_1(1235)$ signal 
and a wide peak centered  
near 1700 $MeV$ are
seen at the total mass spectrum (Fig.5a 
). 
For the $(\pi^+\pi^-\pi^0)$ system
one can see a strong peak from the
$\omega\rightarrow\pi^+\pi^-\pi^0 $ decay
and also the $\eta\rightarrow\pi^+\pi^-\pi^0 $ peak
in Fig.5b, 
as well as 
an accumulation
of events at the mass close to 1300 $MeV$ is seen             
which is close to the $f_1(1285)$
mass. Detailed analisis of this structure is given below.
Concerning the $(\pi^+\pi^-)$ mass spectrum (Fig.5d),
a sharp peak from $K^0 \rightarrow \pi^+\pi^-$ decay is seen 
as well as 
a sharp peak near $780\:MeV$,
the later one is consistent with the $\omega$ mass
and should be attributed to the 
suppressed $\omega\rightarrow\pi^+\pi^-$ decay.
Wide background under the $\omega$ signal originates from 
the $\rho\rightarrow\pi^+\pi^-$ decay. 

\begin{figure}
  \includegraphics[height=.3\textheight]
{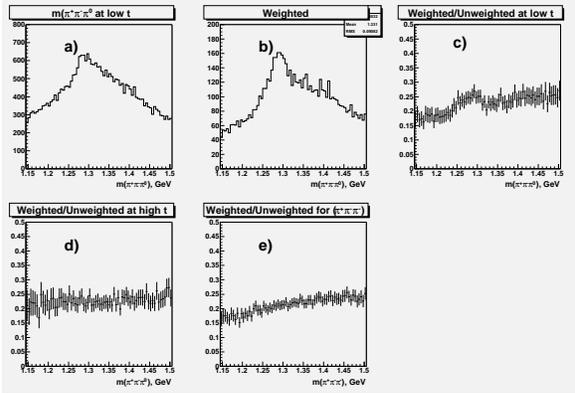}
  \caption{
The three-pion mass spectra for                    
$0.970<m(\pi^+\pi^-)<1.000$ :
a) $m(\pi^+\pi^-\pi^0)$ spectrum at low $|t'|$;
b) like the previous one but weighted;
c) ratio of two previous distributions, weighted/unweighted;
d) ratio of weighted to unweighted mass spectra for $(\pi^+\pi^-\pi^0)$
system at high  $|t'|$;
e) ratio of weighted to unweighted mass spectra for $(\pi^+\pi^-\pi^-)$
system at low  $|t'|$;
}
\end{figure}

It worth mentioning that the $f_1\rightarrow a_0\pi^0$ sample     
originates from the diffractive production. The subsequent processes,
$a_0\leftrightarrow f_0$ mixing and $f_0 \rightarrow \pi^+\pi^-$ decay,
lead to four-pion final state.    
The background 
processes, $\pi^- N \rightarrow (\pi^+\pi^-\pi^-\pi^0) N$, i. e.
production of four 
pions is not a diffractive 
process and it is suppressed. 
This suppression should facilitate the observation. 

To  improve the signal to background ratio,
the following selection criteria have been applied:
a) events at the low momentum transfer, $|t'|<0.04$ were selected;
b) events with a signal detected in the target guard system were rejected;
c) events with $m(\pi^+\pi^-\pi^0)<0.800\:GeV/c^2$ at any combination 
were rejected.
First two cuts tend to select diffractive reaction, 
the third one rejects events with $\omega(780)$ or  $\eta(550)$.

%

Apart from those general cuts, the event selection in different
mass intervals for the $(\pi^+\pi^-\pi^0)$ 
and $(\pi^+\pi^-)$
were tested. The $m(\pi^+\pi^-\pi^0)$ distribution, which  was
obtained with requirements on the two-pion mass 
$0.970<m(\pi^+\pi^-)<1.000\:GeV$, is presented in Fig.6a. 
Clear peak is observed, and its mass is close to the 
$f_1(1285)$ mass. The effect, which arises from the 
application of the angular weight 
$W$ 
to the same event sample, is demonstrated in Fig.6b and 6c. 
A peak at the same mass region is observed
in the ratio of weighted distribution to the unweighted one.
A similar procedure was applied for two another samples,
namely to the  event sample which was selected 
at large $|t|$ and to the $(\pi^+\pi^-\pi^-)$ system at low $|t|$.
The ratios of the weighted to unweighted distributions
are shown in Fig.6d and 6e, 
respectively.
No signal is observed. 

%

   It is also possible to subdivide the event sample at 
low $|t|$ into bins on the three-pion mass and look 
for the mass spectrum of the two-pion system in individual bins.
The mass bin width of $10\:MeV$ was chosen and the $m(\pi^+\pi^-\pi^0)$
interval from $1200$ to $1350$ $MeV$ was subdivided to 15 bins.
The resulting spectrum for the mass bin $(1280,1290)\:MeV$
is shown in Fig.7. 
The $\omega\rightarrow\pi^+\pi^-$ decay
is seen, and another peak with mass close to $985\:MeV$.
A fit by a sum of the Gaussian function for signal and
a background term was performed in the mass interval from 880 
to 1100 $MeV$. 
The product of three-particle phase space by
a quadratic function 
with free coefficients 
was chosen as the background term.
The fit at this bin yields the gaussian mean of $m=983\pm 3\:MeV$
and the gaussian $\sigma=18\pm 4\:MeV$.
The fit $\chi^2/ND=39.8/40$ and the statistical significance
of the gaussial signal is $6.4\:\sigma$.

\begin{figure}
  \includegraphics[height=.3\textheight]
{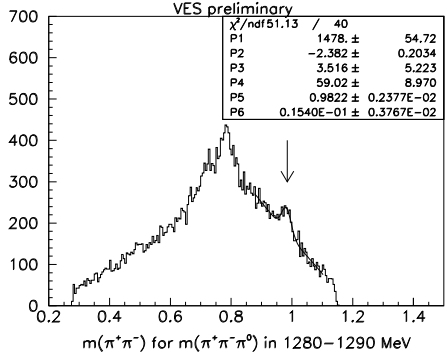}
  \caption{
 $m(\pi^+\pi^-)$, selected combinations with $m(\pi^+\pi^-\pi^0)$ in the mass
interval (1.280,1.290) $GeV/c^2$, 
}
\end{figure}

Similar fitting procedure was applied to all mass bins 
mentioned above with parameters as determined from the central bin. 
The result for the number of signal events in all mass bins
is presented in Fig.8. 
The total number of 
events from decay $f_1 \rightarrow \pi^+\pi^-\pi^0$ in all bins is 
$1572\pm 227$. 
This number of events, taken together with the number of events
in $f_1 \rightarrow \eta\pi^+\pi^-$ channel, gives the 
relative branching ratio. The ratio of the detection efficiencies,
$R = \varepsilon(\pi^+\pi^-\pi^0)/\varepsilon(\eta\pi^+\pi^-)$
was estimated from a Monte-Carlo simulation and taken into account,
$R = 0.95 \pm 0.05$.

The measured dependence of the observed signal
on the $m(\pi^+\pi^-\pi^0)$ can be fitted by a 
Breight-Wigner function, and the result of this fit 
is shown in Fig.8. 
The fitted peak has $m=1288.3 \pm 2.6 \:MeV$
and the width $\Gamma = 21 \pm 4\:MeV$,
which are  
in good agreement with the table values.

\begin{figure}
  \includegraphics[height=.3\textheight]
{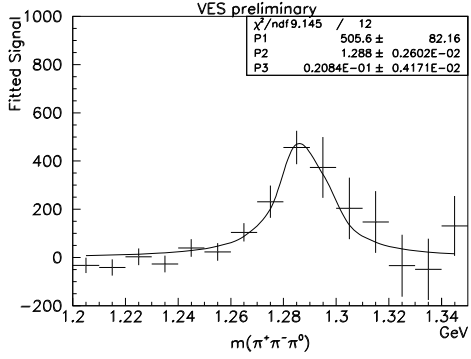}
  \caption{
$m(\pi^+\pi^-\pi^0)$; 
fitted number of signal events as a function of $m(\pi^+\pi^-\pi^0)$;
}
\end{figure}

We tested a presence of a similar signal in charge
mode by means of a similar procedure, i.e.
by subdivision of the event sample into  bins
on the $m(\pi^+\pi^-\pi^-)$ and looking for the 
$m(\pi^+\pi^-)$ spectrum in individual bins.
No signal is observed in the vicinity of the $f_1(1285)$. 

\section{Discussion and conclusions}


One can see  that the signal at $m(\pi^+\pi^-) \sim 985 \: MeV/c^2$ 
is associated with the peak at $m(\pi^+\pi^-\pi^0)=m(f_1(1285))$
having $J^{PC} = 1^{++}$.


All elements of the obseved pattern fit well with predictions based
on the mechanism suggested by Achasov and collaborators 
in 1979 \cite{ref:achasov1979}.


The relative branching ratio is determined 
from the observed number of events in the 
$\eta\pi^+\pi^-$ and $\pi^+\pi^-\pi^0$ channels.
The experimental efficiencies  
for both reactions
are very similar. 
We estimate \\
$\frac{BR(f_1\rightarrow\pi^+\pi^-\pi^0(0.96<m(\pi^+\pi^-)<1.01))}
{BR(f_1\rightarrow\eta\pi^+\pi^-)\cdot
 BR(\eta\rightarrow\gamma\gamma)} = (1.41\pm 0.21 \pm 0.42) \%$; \\
here statistical and systematic errors are indicated.
This relative branching ratio is consistent with estimation
made by Achasov et al. \cite{ref:achasov1981}.


With PDG values for $BR(f_1 \rightarrow \eta\pi  \pi  ) = 0.52 \pm 0.16$
and $BR(\eta \rightarrow \gamma \gamma) = 0.3939 \pm 0.0024$
\cite{ref:PDGf1,ref:PDGeta}
it leads to \\
$BR(f_1\rightarrow\pi^+\pi^-\pi^0(0.96<m(\pi^+\pi^-)<1.01)) =(0.19 \pm 0.09)\%$.



%
%
\section{Acknowledgements}
 This work is supported in part by the Russian Foundation of Basic Research
grants RFBR 07-02-00631 
and by Presidential grant 
NSh 5911.2006.2. 
%
\bibliographystyle{aipproc}   
\IfFileExists{\jobname.bbl}{}
 {\typeout{}
 }

\end{document}